\documentclass{appolb}
\usepackage{epsfig}
\begin{document}
\title{K-mesic nuclei versus eta-mesic  nuclei
\thanks{Presented at International Symposium on "Mesic nuclei", Krak\'{o}w, June 16, 2010}}
\author{ S{\l}awomir Wycech\\
\address{Andrzej So{\l}tan Institute for Nuclear Studies, Ho\.za 69, 00-681 Warsaw,
Poland}}

\maketitle

\begin{abstract}
The nuclear states of $\overline{K}$ and $\eta$ are bound by a
similar mechanism - the excitations of nucleons to $\Lambda(1405)$
and $N^*(1535)$ resonant states.   The observed large differences
in binding energies are understood  in terms of separation of the
involved energies and the resonance positions. The other
experimental findings: broad $\overline{K}$-mesic  and narrow
$\eta$-mesic states are more difficult to understand. A
phenomenological model for $\eta$-N interactions is used  to
explain  the suppression of the $\eta$ absorption in light nuclei.
\end{abstract}
\PACS{13.75.-n, 25.80.-e, 25.40.Ve}

\section{Introduction}
\label{intro}

The possibility of $\eta$-nuclear quasi-bound states  was first
discussed by Haider and Liu  \cite{bha,hai} , when it was realized
that the $\eta$-nucleon interaction is attractive. Nevertheless,
the early ($\pi$,p) experiment looking for these effects was not
conclusive \cite{chrien}. The   reasons  for uncertainties in the
interpretation of those  results is a high background  and
apparently large widths of those states  due to predominantly to
the $(\eta,\pi)$ conversion. Recent detailed calculations
\cite{hir09} indicate the states to be wide and difficult to
extract experimentally.

The view that the widths of nuclear $\eta$ states are large is
widespread. Calculations \cite{hai} and  especially those based on
chiral models  \cite{hir09},\cite{gar02},  predict large widths.
If the quasi-bound states exist, one  expects these to be
narrower, and easier to detect  in the  few-nucleon systems.
Indeed, an indirect evidence  was suggested by Wilkin \cite{wil},
who interpreted a rapid slope of the $ pd \rightarrow \eta ^{3}$He
low energy amplitude as a signal of a quasi-bound state. Later, a
very strong three-body $d\eta $ correlations were found in
measurements of the $ np \rightarrow d \eta$ cross section in the
threshold region \cite{cald}. Calculations indicate that the
$d\eta$ system forms a virtual state \cite{wyc01}. The status of
$\eta ^{3}$He state is still unsettled.

Some of these results  have been  superseded  by new experiments:

$\bullet$ The final state interactions in the $\eta ^{3}$He system
obtained by COSY-ANKE determines  large scattering length $ A(\eta
^{3}He) = \pm 10.7 (\pm 0.8 ^{(+0.1)}_{(-0.5)})  +i 1.5(\pm 2.6
^{(+1.0)}_{(-0.9)}) $ \cite{mer07}.

$\bullet$ The final state interactions in the $\eta ^{4}$He system
obtained by GEM  determines  scattering length $ A(\eta ^4He) =
\pm 3.1 (5)  +i0.0(5)$ fm  \cite{bud09a}.

$\bullet$ The reaction $ ^{27}Al +p \rightarrow  ^3He +p + \pi^- +
X$ studied by  COSY-GEM  attributed $X$ to the state of
$\eta^{25}Mg $ nucleus. The energy $E_B \approx -13 - i 5 $ MeV
was found  \cite{bud09}.

These findings require rather weak absorption  and contradict many
theoretical calculations based on the single channel $\eta$-N or
multiple nucleon absorption  models. In this paper the latest
Helsinki $K$-matrix model incorporating the
$\eta$-N,$\pi$-N,$\gamma $-N channels is used \cite{gw05}.  It
offers  two characteristic features : large scattering length $
a_{\eta N}=0.91(6)+i0.27(2) fm $ and a rapid decrease of the
absorptive scattering amplitude in the subthreshold region.

\emph{The essential point of   this  work } is the observation
that  in the few nucleon systems the relevant $\eta$-N, scattering
amplitude involves subthreshold energies in the $\eta$-N  center
of mass system. The quantity of interest is
\begin{equation}
\label{f1} T_{\eta N}( E_{cm}= - E_N - E_{\eta} - E_{recoil})
\end{equation}
where $E_N$ and $ E_{\eta}$  are binding energies  of  a nucleon
and the  meson. To obtain the  center of mass energy the recoil
energy of the meson-nucleon pair with respect to the residual
system has to be subtracted  in the argument of  $ T_{\eta N}$.
Such amplitudes are the standard input in the three-body Faddeev
equations for the bound states and low energy scattering. There
are other situations where   the subthreshold amplitudes are
appropriate: the interactions at nuclear surfaces and the tightly
bound residual systems. To certain degree these situations are met
in the states of $\eta$ mesons bound to light nuclei.

The absorptive part   Im  $T_{\eta N}( E_{cm})$  determines the
dominant part of the level widths.  It is proportional to the
phase space in the $\pi$-N decay channel given by the center of
mass energy $ E_{cm}$.  Thus the argument $E_{cm}$ given in
Eq.(\ref{f1}) is proper, at least in the Im $T_{\eta N}( E_{cm})$,
for a  much wider class of systems.

Extension to subthreshold energies reduces absorption in the
$\eta$ systems. On the other hand there are additional effects
which enhance the level width. One is the two nucleon absorption
of the meson. In the eta case there is an experimental check on
the related rate coming from the $\eta$ formation in the two
nucleon collisions. This rate is  low  \cite{kul98}.  Another
effect is the multiple scattering in the $\pi$-N decay channel. It
 goes beyond the optical potential approach and it is known to be
significant on the K-mesic  nuclei \cite{she07,wyc09}. These
questions are discussed in the main body of this paper with the
special reference to the three recent experimental results.

\emph{In conclusion}: (1) there are models of $\eta$-N
interactions which generate fairly narrow  $\eta$ states in light
nuclei. (2) Some systems in particular the   $\eta ^3$He are
difficult to calculate
 precisely. Due to the apparently  large scattering length all secondary
effects become significant. (3) The nuclear states of
$\overline{K}$ mesons indicate a need for explicit description of
the decay channels. This question should be approached also in the
$\eta$ meson case.

\section{  Subthreshold eta-nucleon scattering amplitude }
\label{sec2}

The latest Helsinki $K$-matrix model incorporates  the $\eta$-N,
$\pi$-N, $\gamma $-N channels. It is presented in refs.
\cite{gw05} and  only the main points are indicated here.  The
scattering data are parameterized in terms of a phenomenological
$K_{i,j}$ -  a matrix in the channel  indices $i,j$. Next, linear
equations for the scattering  matrix  T
\begin{equation}
\label{k1}
 T_{i,j}=  K_{i,j} +  i  \Sigma_m~ K_{i,m}~Q_{m}~T_{m,j}
\end{equation}
are solved   with $ Q_{m} $ being the diagonal matrix of the CM
momenta in each channel. The energy region of interest for the few
body eta physics spans from about 40  MeV below the eta-nucleon
threshold to some 20 MeV above it. This region is dominated, in
both channels, by the $N^*(1535)$. The model used here supposes
this state to be determined by some short range interactions.
Next, this state is coupled to the channel states which change its
properties. However, to obtain a better restriction of the
parameters the region of the $K$ matrix description is extended to
about 200 MeV below and above the threshold. So, the higher
N(1650) resonance is also included.

The K matrix is parameterized as
\begin{equation}
\label{k2}
 K_{i,j}=\Sigma ~
\frac{\sqrt{\gamma_{i} \gamma_{j}}}{E_o-E} + B_{i,j}
\end{equation}
where the sum extends over two resonances represented by the  pole
terms. The $\gamma_{i}$ couple these  to the channels. The
additional background matrix $B_{ij}$ describes other forms of the
interactions. These change the bare resonance energies $E_o $ to
those observed in the scattering experiments. The free parameters
$\gamma_{\pi}, E_o , B_{ij}$ obtained by the best fit to the data
may be found in refs. \cite{gw05}.  One obtains several sets
depending on the choice of the input data. The best result for the
elastic $\eta$-N amplitude is plotted in Fig.1. It is only
marginally better than other possibilities. The main difference
happens in the value of $ B_{\eta N,\eta N}$ reflected in
different strength of the spike in Re $T_{\eta,N}$ at the
threshold i.e. the scattering length. On the other hand the
absorptive part Im $a_{\eta,N}$ stays close to the "canonical"
value of 0.27(2) fm. The rapid decrease of Im $T_{\eta,N}$  occurs
in all solutions.

\begin{figure}[ht]

\begin{center}
\includegraphics[width=0.8\textwidth]{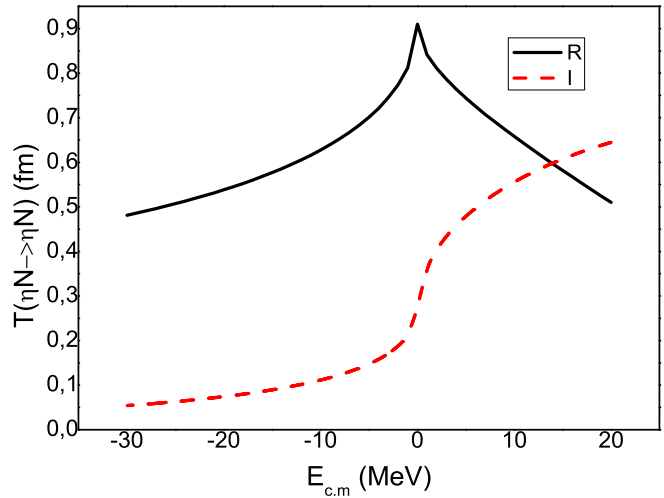}
\caption[F1] {The elastic $\eta$-N scattering amplitude  plotted
against the C.M. kinetic energy. Real part - continuous line,
absorptive part - dashed line. This amplitude is well represented
by the effective range expansion
$T_{\eta,N}^{-1}+iq_{\eta}=1/a+r_0/2 q^2_{\eta}+s q^4_{\eta}$,
 -- with $ q_{\eta}$ being the momentum in the $\eta N$
 center-of-mass and
  $a= 0.91(6)+i0.27(2),  r_0= -1.33(15)-i0.30(2), s= -0.15(1)-i0.04(1)$. All in $fermi$
units.} \label{fig1}
\end{center}
\end{figure}

\section{ The absorption of $\eta$-mesons in light nuclei}
\label{sec2}
\subsection{ Helium }

In this section the $\eta$-He scattering lengths are calculated.
The energy argument  entering eq.(\ref{f1}) in $^4$He is
determined by the nucleon separation $E_{N}\approx - 21$ MeV and
$E_{\eta}=0$.  The meson-nucleon  recoil energy is a function of
total pair momentum $P$ and the momentum distribution $ f(P)$  is
calculable with the meson and nucleon wave functions
\begin{equation}
\label{h1}
 f(P)= \int d \textbf{r}
 ~\phi_N(r)\phi_{\eta}(r)~\exp(i\textbf{Pr}),
\end{equation}
where $\textbf{r}$ is the coordinate relative to the  tritium (or
$^3$He) core.  The  $f(P)^2$  is peaked around  the average
momentum and the average recoil is given by  $E_{recoil}=
<P^2>/(2M_{\eta N, R}) $ where $M_{\eta N, R}$ is the
corresponding reduced mass. For low energy mesons  one has
$E_{\eta} \approx 0, $  $\phi_{\eta}(r)\approx const$ and $
E_{recoil}= 16 $ MeV. The average subthreshold energy is
$E_{cm}\approx -37 $ MeV and  the amplitude becomes
$T_{\eta,N}(E_{cm}) = 0.45 +i 0.051$. For the other isotope $^3$He
the corresponding values are $E_{N}\approx - 7$, $ E_{recoil}= 12
$, $E_{cm} \approx -19 $ MeV and $T_{\eta,N}(E_{cm}) = 0.54 +i
0.077$ fm. The width of momentum distribution is about 10 MeV and
inspection of Fig.1 indicates that the $T_{\eta,N}(E_{cm})$ is
fairly constant in the regions of interest.

These effective  scattering matrices $T_{\eta,N}(E_{cm})$  are now
used to calculate the $A(\eta,He)$ scattering lengths. The $
\eta$-He multiple scattering series is summed according to the
prescription of ref.\cite{gwn95} equivalent to  the calculation in
terms of the optical potential. The latter is given in the
standard way
\begin{equation}
\label{pot1}
  V_{N}(r) = - \frac{2 \pi}{\mu_{\eta N}} ~ T_{\eta N}(E_{cm}) ~  \rho(r) ,
\end{equation}
where  $\rho $ is the nuclear density, $\mu $ is the $\eta$-N
reduced mass and the index $N$ on $ V_N$ indicates the single
nucleon origin of this potential.

$\bullet$   For $ ^4$He this input generates  $A(\eta,^4He)= -2.90
+ i 0.35$ fm, which corresponds to a quasi-bound state of energy $
E \approx -6 $ MeV and width $\Gamma \approx 3.0$ meV. This
compares well with the experimental $A(\eta,^4He)= \pm3.1(5)+ i
0.0(5) $fm, \cite{bud09a}.

$\bullet$   In  $ ^3$He  both  the theoretical   and experimental
situation is uncertain as apparently the singularity in the  $\eta
^3$He scattering matrix is close  to the threshold. On the
experimental side  COSY-11 \cite{smy07} obtains $|A(\eta,^3He)|=
4.3(5)$ fm. This result  is consistent with the  phenomenological
$ A(\eta ^{3}He) = 4.24(29)+i 0.72(81) $ fm based on older data
\cite{gw03}. In the latter case, the inclusion of $(\pi ,\eta) $
data  allowed  to establish the sign of the real part which
signals a virtual state. These lengths indicate that  the related
singularity of the $T(\eta ^{3}He)$ matrix is located in the
complex energy plane  some 1.5-2 MeV away from the threshold.

On the other hand the  COSY-ANKE solution is $A_{\eta,N}= \pm 10.7
(\pm 0.8 ^{(+0.1)}_{(-0.5)}) + i 1.5(\pm 2.6 ^{(+1.0)}_{(-0.9)}),
$ \cite{mer07}. This   value indicate the pole to be  only 0.3 MeV
away from the threshold. To obtain it one needs a strong
suppression of the meson absorption.

The photo-production result  \cite{pfe04} indicates  a quasi-bound
state of energy $ E = -4.4(4.2 )- i 12.8(3.1)$ which corresponds
to a much smaller scattering length.

$\bullet\bullet$ Calculations of large scattering lengths are
unstable. Indeed, with equation (\ref{pot1}) and the effective
$T_{\eta,N}(E_{cm}) = 0.54 +i 0.077$ obtained from fig1. one
obtains a large length $A_{\eta,N}= -6.2+i 2.8$ fm. However, a
simple correction introduced  to the multiple scattering series,
the  replacement of $A^2$ by $A(A-1)$  in the double scattering
term, changes the result to $A_{\eta,N}= 7.3 +i 7.7$ fm. This
shows the outcome to be   unstable against  second order effects.
 A better calculating techniques are also required.

\subsection{ Magnesium }

In this section a crude estimate of the $\eta$ mesic  level in
$^{25}$Mg is given. With the  characteristic value  of Re
$T_{\eta,N}(E_{cm}) = 0.52 $ fm the optical potential  generates
the $\eta$  binding of about  18 MeV. The radius m.s. of the meson
density distribution becomes 3.2 fm comparable to the charge
density radius of 3.11 fm \cite{VRI87}.  Thus  the meson is mostly
located  in the region beyond the half density radius which in
this nucleus is 2.76 fm. The nucleon separation energies are
determined mostly by the upper single particle levels  with an
average  about 15-20 MeV. Following Fig.1 one obtains
$T_{\eta,N}(E_{cm}) \approx 0.52 +i 0.07 $ fm. This generates
narrow width $ \Gamma \approx 6.0$ MeV.

One concludes that the suppression of the level widths can be
understood at least on the semi-quantitative level. However, the
binding offered by the K-matrix model seems  to be excessive. In
addition  a number of higher order effects must be included:

1) Two nucleon $\eta$NN  capture. There is an experimental check
on this effect to be discussed below. It adds some 1-2 MeV to the
width.

2) Interactions  in the decay channel

3) Nuclear medium effects  change $T_{\eta,N}(E_{cm})$. In the
light nuclei these are hard to calculate.

\section{ Other absorption modes }

The $\eta$ meson lifetime in a nucleus is determined by the basic
reactions
\begin{equation}
\label{d1} \eta N \rightarrow \pi N
\end{equation}
\begin{equation}
\label{d11} \eta N \rightarrow \pi \pi N
\end{equation}
\begin{equation}
\label{d2}
 \eta (NN)^0 \rightarrow N N
\end{equation}
\begin{equation}
\label{d3} \eta (NN)^1 \rightarrow N N.
\end{equation}

Where the superfix denotes the spin of  NN pairs. The first
process is known  fairly  well, the second one is usually included
into absorptive $T_{\eta,N}$ amplitude due to the two pion decay
of the $N^*(1535)$.

The other two reactions (\ref{d2}) and (\ref{d3}) correspond to
$\eta$ absorption on two correlated NN pairs in either the spin
singlet or spin triplet states. A  phenomenological evaluation of
the rates is possible as the cross sections for
\begin{equation}
\label{r2}
  p p \rightarrow p p \eta
\end{equation}
\begin{equation}
\label{r3d}
 p n  \rightarrow d \eta
\end{equation}
\begin{equation}
\label{r3}
 p n  \rightarrow  p n \eta
\end{equation}
have been  measured in the close to threshold region. The analysis
based on the detailed balance corrected for final state
interaction has been performed in ref. \cite{kul98}. Absorptive
potential of the $ \rho(r)^2 $ profile with a weak strength Im $
W_{NN}( r=0) = 3.2 $ MeV was obtained.

An additional absorption mode exists if the decay channel is
described explicitly. It has been studied in terms of Faddeev
equations used to  calculate  the $\bar{\textrm{K}}$ NN
quasi-bound state energy \cite{she07}. An explicit treatment of
the multiple scattering in the decay channel generates an
additional binding and enlarges the width of the state. Similar
effects are found in a variational calculations of the
 $\bar{\textrm{K}}$- few- N levels \cite{wyc09}.

\subsection{Interactions in the decay channels}

A simple model of the  $\bar{\textrm{K}}$ interacting with two
fixed nucleons is used to explain the effect (a finer presentation
 may be found in ref.\cite{wyc09}). Consider scattering of the
meson bound to two nucleons fixed at a separation $\textbf{r}$.
Let the amplitudes of the meson  at each nucleon be $ \psi_1,
\psi_2$. The meson bounces off each nucleon and the multiple
scattering equations are
\begin{equation}
\label{a1}
 \psi_1 + t~G ~\psi_2 = 0, ~~~~  \psi_2+t~G~\psi_1 = 0,
  \end{equation}
where $t$ is the meson-nucleon scattering matrix and $G$ is the
propagator for  the meson passing from one to the other nucleon
\begin{equation}
\label{a2}
 G = G(p,r)=
\frac{1}{ r} ~ \exp(ipr)
  \end{equation}
One needs to regularize $G$ at  short ranges but for simplicity of
the presentation this is suppressed.  The consistency between the
scattering and the bound state requires vanishing of the
determinant
\begin{equation}
\label{a3} D=  1-( t~G)^2 = (1+tG)(1-tG)= 0.
 \end{equation}
This condition  determines the \emph{ complex }eigen-momentum
$p(r)$ which gives the energy and the width of the meson +
fixed-NN system.

If the $\bar{\textrm{K}}$N  interaction is dominated by a
resonance below the threshold, such as $\Lambda(1405)$, then $ t =
\gamma^2/(E-E_o +i\Gamma/2)$, where
 $\gamma$ is a coupling constant and
  $E_o-i\Gamma/2$ is the $\Lambda(1405)$ complex  energy.
  The solution of  eigenvalue equation,   $ 1+ tG= 0 $,  takes  the form
\begin{equation}
\label{a4} E  =   E_o -i\Gamma/2 - \gamma^2 G(r,p).
\end{equation}
The     solution $ E(r)  \equiv E_B(r)  - i \Gamma(r)/2$ depends
on the N-N separation  $r$.   Since Re $ G(r,p)$
 close to the resonance is positive,  the
binding  of $\bar{\textrm{K}}$ to fixed two nucleons  is stronger
than the $\bar{\textrm{K}}$  binding to one nucleon, $\mid
E_B(r)\mid
> \mid E_o\mid$. Increasing the separation $r\rightarrow\infty$
one obtains $G \rightarrow 0$ and $ E(r) \rightarrow  E_o $,~ i.e.
the $\bar{\textrm{K}}$ meson becomes bound to one of the nucleons.
In the same limit the lifetime of $\bar{\textrm{K}}$ becomes equal
to the lifetime of $\Lambda(1405)$. Hence, the separation energy
is understood here as the energy needed to split the
$\bar{\textrm{K}}$-N-N system into the $\Lambda(1405)$-N system.
The last  term in Eq.(\ref{a4})  constitutes a potential
contracting the two nucleons. It is very strong and leads to large
50-100 MeV bindings of the system and the widths in the range of
40-80 MeV. The next step in the calculation (not presented ) is to
allow the nucleon degrees of freedom and use these results in a
variational procedure.

The decay channel $ \Sigma\pi$  coupled to the basic
$\bar{\textrm{K}}$ N channel may be  introduced  explicitly. The
wave function has two components, one related to  the
$\bar{\textrm{K}}$ N the other  to   the  $ \Sigma\pi$ channel.
The scattering amplitudes are  two dimensional vectors $ \psi_i
\rightarrow [\psi_i^K,~ \psi_i^\pi $] at each nucleon. Now $t$
becomes a matrix in two channel indices $ t_{a,b} = \gamma_a
\gamma_b~/~ (E - E_o+ i \Gamma/2)$, where $a,b = K,\pi$, and below
the threshold $\Gamma/2= (\gamma_{\pi})^2 ~ p_\pi$. The multiple
scattering equations  are  changed accordingly and the binding
energy
\begin{equation}
\label{t5} \emph{Re } E =  E_o  - (\gamma_K)^2\frac{cos(p_Rr)}{r}
\exp(-p_Ir) - (\gamma_{\pi})^2 \frac{cos(p_\pi r)}{r}
\end{equation}
becomes larger than the binding of the resonance  but the
collisions in the  decay channel induce oscillations.
 This
oscillatory behavior is also seen in the width of the system
\begin{equation}
\label{t6}  \emph{Im}~E = - (\gamma_{\pi})^2 ~ p_\pi~[
 1+\frac{sin(p_{\pi}r)}{p_{\pi}r}] -  (\gamma_K)^2\frac{sin(p_Rr)}{r}\exp(-p_Ir).
\end{equation}
The contribution from multiple scattering in the decay channel is
sizable in general but it oscillates and may under some conditions
reduce the total width. That is an effect of interference in  the
decay channel. In the $\bar{\textrm{K}}$ NN case the scattering in
the decay channel turns out to be constructive  and leads to about
25$\%$ stronger binding and larger widths.

Unfortunately,  in the $\eta$ meson case this method cannot be
used as the $N^*(1535)$ is located above the $\eta$-N threshold.
The solutions given above exist  for  N-N distances less than  a
critical value $R_c$. In the case of $^4$He  one has $R_c\approx
1.5 $fm and the variational method of ref.\cite{wyc09} seems
applicable. In the most interesting  $^3$He case it is not. Other
methods should be tried as the effects might be sizable.

\vspace{0.5cm}

\noindent \emph{Acknowledgements}. This work has been  supported
by the Hadron 2 European Project   LEANNIS .

\end{document}